\documentstyle[12pt,aaspp4]{article}

\def\hm#1#2{\ifmmode{ #1^{\rm h} #2^{\rm m}}\else{ $ #1^{\rm h} #2^{\rm m}$}\fi}
\def\hms#1#2#3{\ifmmode{#1^{\rm h} #2^{\rm m} #3^{\rm s}} \else{$ #1^{\rm h} #2^{\rm m} #3^{\rm s}$}\fi}
\newcommand\kms{\hbox{${\rm \ km}\thinspace{\rm s}^{-1}$}}

\begin{document}
\slugcomment{To appear in {\it The Astrophysical Journal}, Vol. 483}

\title{The ROSAT-HRI X-Ray Survey of the Cygnus Loop}

\author{N.~A.~Levenson\altaffilmark{1}, J.~R.~Graham\altaffilmark{1}, 
B.~Aschenbach\altaffilmark{2}, W.~P.~Blair\altaffilmark{3},
W.~Brinkmann\altaffilmark{2}
J.-U.~Busser\altaffilmark{2},R.~Egger\altaffilmark{2},
R.~A.~Fesen\altaffilmark{4}, J.~J.~Hester\altaffilmark{5},
S.~M.~Kahn\altaffilmark{6}, R.~I.~Klein\altaffilmark{1,}\altaffilmark{7}, 
C.~F.~McKee\altaffilmark{1,}\altaffilmark{8}, R.~Petre\altaffilmark{9},
R.~Pisarski\altaffilmark{9}, J.~C.~Raymond\altaffilmark{10}, 
and S.~L.~Snowden\altaffilmark{9}
}

\altaffiltext{1}{Department of Astronomy, University of California, Berkeley, CA 94720}
\altaffiltext{2}{Max--Planck--Institut f\"ur Extraterrestrische Physik,
 Giessenbachstrasse, D-85740 Garching, Germany}
\altaffiltext{3}{The Johns Hopkins University,  Department of Physics and
Astronomy, Baltimore, MD 21218}
\altaffiltext{4}{Department of Physics and Astronomy, Dartmouth College,
Hanover, NH 03755}
\altaffiltext{5}{Department of Physics and Astronomy, Arizona State University,
Tempe, AZ 85287}
\altaffiltext{6}{Departments of Astronomy and Physics, Columbia University, 
New York, New York 10027}
\altaffiltext{7}{Lawrence Livermore National Laboratory, Livermore, CA 94550}
\altaffiltext{8}{Department of Physics, University of California, Berkeley, CA 94720}
\altaffiltext{9}{Goddard Space Flight Center, Greenbelt, MD 20771}
\altaffiltext{10}{Center for Astrophysics, 60 Garden St., Cambridge, MA 02138}

\begin{abstract}

We describe and  report progress on the joint U.S. and German campaign to map the
X-ray emission from the 
entire Cygnus Loop supernova remnant with the ROSAT High Resolution
Imager.  The Cygnus Loop is the prototype for a
supernova remnant that is dominated by interactions with the
interstellar medium
 and  supplies fundamental physical information on this
basic mechanism for shaping the interstellar medium.  The global view
that  these high-resolution ($FWHM \sim 10\arcsec$) observations provide
 emphasizes the inhomogeneity of the interstellar medium and the
pivotal nature of cloud-blast wave interactions in determining the
X-ray morphology of the supernova remnant.  While investigating the
details of the evolution of the blast wave, we also describe the
interstellar medium in the vicinity of the Cygnus Loop, which the
progenitor star has processed.  Although we do not expect the
 X-ray observations
to be complete until September 1997, the incomplete data  combined with
deep H$\alpha$ images provide definitive
evidence that the Cygnus Loop was formed by an explosion within a
preexisting cavity.

\end{abstract}


\section{Introduction}

Supernova explosions are one of the most basic processes that define the
physical and chemical state of the interstellar medium (ISM).  The
interaction between supernova remnants (SNRs) and the ISM is complex and
symbiotic. X-ray observations are important in the study of SNRs because
thermal emission from hot material ($kT \sim 0.1$--1~keV) is the primary tracer
of gas that the supernova remnant blast wave heats and accelerates.  
High resolution X-ray
observations are important for studying SNR gas dynamics for three
reasons.  First, the
length scale, $\ell$, for cooling and recombination behind a radiative
shock is small: $\ell \approx 2.5 \times 10^{17} v_{s7}^4/n_0 ~{\rm cm}$,
where $n_0$ is the preshock number density, and $v_{s7}$ is the shock
velocity in units of $100~{\rm km~s^{-1}}$ (McKee 1987\markcite{McK87}). 
Second, the preshock medium
is not uniform, having clouds  and substructure ranging over scales from
tens of pc and downward.  Third, gas-dynamical instabilities are
important, and multi-dimensional hydrodynamic calculations
  predict that fully developed turbulence
occurs when the blast wave collides with an interstellar cloud  (Klein, McKee,
\&\ Colella 1994\markcite{Klein94};
Klein, McKee, \& Woods 1995\markcite{Klein95}). Consequently, structure 
is present down to the
smallest scales that can be resolved computationally. The improved
ROSAT-HRI sensitivity and spatial resolution compared to {\it Einstein}
make it invaluable for exploring cloud-blast wave interactions.  The
ROSAT-HRI combines high 0.1-2.4~keV sensitivity (approximately
three times
better than the {\it Einstein} HRI; Zombeck et al. 1990\markcite{Zom90}), high
spatial resolution ($FWHM \approx 6\arcsec$ on axis),
and high contrast imaging (the half-energy radius is
 $3\arcsec$ on axis, with
scattering less than a few percent; Aschenbach 1988\markcite{Asch88})
to provide a powerful tool for investigating cloud-blast wave interactions.

The region of the Milky Way towards Cygnus is viewed along the
Carina-Cygnus arm and exhibits a complex array of bright and dark
clouds on the Palomar Sky Survey.  Most remarkable are the arcs and
loops of faint H$\alpha$ nebulosity, the smallest and brightest of which is
the Cygnus Loop ($\ell = 74.0$, $b = -8.6$).
The Cygnus Loop is the prototype middle-aged SNR and
is a particularly suitable candidate for a case study being  nearby (770
pc; Minkowski 1958\markcite{Mink58}) and bright with low foreground 
extinction [$E(B-V) = 0.1$],
thus ensuring unrivaled signal-to-noise and spatial resolution. From an
observational perspective the Cygnus Loop is unique in terms of X-ray
brightness, shock front resolution ($1" \equiv 1.1 \times 10^{16}~{\rm cm}$),
and interaction with a range of interstellar conditions. The Cygnus Loop is
also at sufficiently high galactic 
latitude that confusion with background emission
from the plane of the galaxy is minimized. Unlike other well-known younger
SNRs such as Cas A, Tycho, and Kepler, the structure of the Cygnus Loop
is dominated by its interaction with its environment. The large
diameter of the Cygnus Loop ($3^\circ \equiv 40$~pc) assures that it is
interacting with different components of the ISM, from molecular gas on
the northwest edge (Scoville et al.  1977\markcite{Sco77}), to diffuse atomic gas 
along the  northeast and eastern limbs 
(DeNoyer 1975\markcite{DeN75}; Hester, Raymond, \&\ Blair 1994\markcite{HesRB94}), 
and low density, hot, ionized gas to the south (Ku et al. 1984\markcite{Ku84}).
Previous studies of the Cygnus Loop have focused on individual
regions selected for their brightness or morphological
peculiarity (e.g.  Hester \&\ Cox 1986\markcite{HesCox86}, 
Hester et al. 1994\markcite{HesRB94}, Fesen,
Kwitter \& Downes 1992\markcite{Fes92}, Graham et al. 1995\markcite{Gra95},
 and Levenson et al. 1996\markcite{Lev96}).
The goal of this high-resolution investigation 
(Graham \&\ Aschenbach 1996\markcite{Gra96}) is to make an unbiased survey and thereby
establish globally  the state of the medium into which the blast
wave expands.  
A coherent picture   emerges from our preliminary results
(\S 2).  We present a unified interpretation of the environment and
development of the Cygnus Loop (\S 3), constrain the global models
that are relevant (\S 4), and summarize the conclusions in \S 5.

\section{Observational Strategy and Data Reduction}

Our primary goal is to map the entire Cygnus Loop
 remnant at the highest spatial resolution
that can be attained with the ROSAT-HRI.  Achieving this ambition is
not simple, because the spatial resolution is limited by the intrinsic
instrumental properties and photon counting statistics.  The on-axis
resolution is about $6\arcsec$, but the off-axis imaging performance is
dominated by telescope aberrations, which increase with increasing
field angle.  At $20\arcmin$, the edge of the useful field of view of
the HRI, the resolution has degraded to about $30\arcsec$. The large range
of surface brightness allows us the luxury of mapping the brightest
regions at full HRI resolution, but forces us to seek a compromise
between effective resolution and observing time in the dimmest
regions.  To give a concrete example, the surface brightness observed
with the ROSAT HRI yields a count rate  of
$3\times10^{-5}$---$1\times10^{-3}~{\rm s}^{-1}$
in a $5\arcsec\times 5\arcsec$ pixel.  For the brightest
regions along the shell (within a factor of two  of the highest surface
brightness) a $15~$ks exposure will net 15 counts and 0.4 background
counts.  It would take longer than $450$~ks to 
obtain a similar number of counts
in the faint interior. We therefore adopted the following observing
strategy.  We use the full $6\arcsec$ resolution to observe the
brightest regions, which restricts the effective field of view to
$8\arcmin$ radius.  Each of these 24 pointings requires 15--20 ks to
obtain 15 counts per pixel.  In a comparable exposure time, the
moderate surface brightness interior regions yield about 10 counts per
$12\arcsec \times 12\arcsec$ pixel.  At this resolution, the effective HRI field
of view has a $12\arcmin$ radius.  We have observed these nine fields
for 20--25~ks.  Observing the low surface brightness regions, we obtain
10 counts per cell only when using a $15\arcsec$ or larger pixels, for
which the full HRI field of view is suitable.  Each of these 10 fields
requires a 35~ks exposure.  Finally, one very low surface brightness
region along the western edge of the ``breakout'' requires a 50~ks
exposure to obtain reliable counting statistics.  The locations of the
HRI pointings are shown on the {\it Einstein} IPC map (Fig. 1).

A few fields were  observed during the initial operation of ROSAT, and
a concerted effort to make a complete map began during AO5, in October 1994.
  Because of the
large total observing time  this program requires,
it is a collaborative project between U.S. and German investigators.
So far 771~ks out of a total 1066~ks of data have been
obtained and we expect completion in AO7, by September 1997.
The complete mosaic will consist of 47, fields of which 37 have been 
observed at least in part.  We present processed data for 26 fields,
of which 15  are complete.  The target names, field centers, and 
exposure times are listed in Table 1, and  the  
observations that are presented here appear first.
Column 4 contains the live time, which is the useful portion of the
observations, and column 5 lists the remainder of the requested time.

The data were processed and combined into a single image using software
designed to analyze extended sources and
the diffuse background (Snowden et al. 1994\markcite{Sno94}).  Over such large 
spatial scales, each observation is a distinct
projection of the sky onto the plane of the detector, and these must be
re-mapped to a common projection.  For each observation,
using only the accepted time intervals listed in the raw data files,
three images are created.  The first map contains the total counts, the
second map is a model of particle background, and the third is an exposure
map, which includes the variations in the detector quantum efficiency.
These images have $5\arcsec$ pixels, matching the instrument's angular resolution.
These maps from individual fields are summed on a pixel-by-pixel basis
into three mosaics of the entire Cygnus Loop field.
For the initial analysis we present here, we use a uniform pixel scale
and field of view from each observation.
 Only the central $17\arcmin$ of each
observation  in order to avoid the significant degradation of
the point spread function and decreased efficiency at large off-axis
angles, except for two fields where a radius of $20\arcmin$ was used
to avoid a gap in the resulting image.  
 The count rate map (Fig. 2) has $15\arcsec$ pixels and is the difference of
the total counts and the total background, divided by the net
exposure (Fig. 3).  
Background subtraction is not a problem for the bright and
moderately bright fields, but the background does become comparable with
the source surface brightness for the lower surface brightness regions.
The extent of the observations is demonstrated in Figure 4, where the 
individual image boundaries are drawn on the count rate map.

To display the high resolution information contained in the
mosaic, we also show three smaller fields that have been binned with $6\arcsec$
pixels (Figs. 5a, 6a, and 7a).  
For comparison, we present aligned optical images of
the same regions taken through a narrow H$\alpha$ 
filter (Figs. 5b, 6b, and 7b).  These were
obtained using the Prime Focus Corrector on the 0.8 m telescope 
at McDonald Observatory and are described in Levenson et al. (1997\markcite{Lev97}).
After basic reduction consisting of bias subtraction and flat-fielding,
individual images were re-mapped and averaged using the same
technique that was applied to the X-ray observations.  Figures
5c, 6c, and 7c are color versions of these combined data sets to show
the relative spatial distributions directly.

\section{Results}

The X-ray morphology of a SNR is a primary diagnostic of the medium
into which the blast wave propagates. At first sight the X-ray
morphology depicted in the ROSAT-HRI mosaic presents a bewilderingly
complex view. This shows {\it prima facie}  that the Cygnus Loop is
dominated by its interaction with an inhomogeneous environment. In
contrast, we note also that the Cygnus Loop appears nearly circular
except for the breakout region to the south, which suggests that
the SNR expands in a uniform medium. Any model for the
Cygnus Loop must resolve this apparent paradox.

There is considerable large- and small-scale azimuthal variation around
the perimeter of the Cygnus Loop and significant structure projected across
its face. X-ray emission is bright near the rim, but the brightest
peaks all lie interior to the circle that circumscribes the X-ray
edge. The two most prominent X-ray regions, NGC 6960 on the western
edge and NGC 6992 to the northeast, lie at the largest projected distance
inside this circle. These regions contain  complex networks of X-ray
filaments and high surface brightness clumps. The emission here is
strongly limb brightened. Figures 8 and 9  show radial cuts of the emission
from these regions. The limb-brightening (i.e. the ratio of brightness
at center to edge) is about a factor of 6 in the northeast (Fig. 8) and 
 30 in the west (Fig. 9).  
The adiabatic expansion of a supernova blast wave in a uniform
medium results in some degree of 
limb brightened X-ray emission but is inadequate to account for these
profiles.   We have calculated the X-ray
surface brightness profile (Fig. 10) using the Sedov density 
profile (Sedov 1993\markcite{Sedov93}),
taking X-ray volume emissivity to be proportional to density squared,
 and assuming that it is independent of temperature, 
which is reasonable for the energy coverage of our observations 
(Hamilton, Sarazin, \&\ Chevalier 1983\markcite{Ham83}).
The maximum limb-brightening is 3.3 and occurs near the edge of the SNR.
The clear message from this comparison is that the observed
X-ray profiles in these
bright regions of the Cygnus Loop have peaks in the wrong place
and too great in contrast to be the result of an adiabatic blast wave
propagating in a uniform medium.  
Thermal evaporation does not 
explain the X-ray profiles either, as the evaporative
solutions of White \&\ Long (1991\markcite{WhL91}) demonstrate (Fig. 10; see also
Cowie, McKee, \&\ Ostriker 1981\markcite{Cow81}).
 The effect of thermal conduction is to
provide more mass farther behind the blast wave, so these solutions
always exhibit less extreme limb brightening than the adiabatic case. 
{\it The Cygnus Loop is neither a Sedov remnant nor 
a Sedov remnant modified by thermal evaporation.}

\subsection{X-ray and Optical Correlations}

Optical line emission alone reveals some details of shock structure.
 For example, faint
H$\alpha$ filaments trace the progress of the blast wave in neutral
gas, [O~III] emission reveals where postshock gas first starts to cool
and recombine, and [S~II] and bright H$\alpha$ together show where complete
cooling and recombination zones have developed.
Utilizing X-ray and optical information together, however, provides a
more complete understanding of the structure and development of supernova
remnant shocks.

The observed correlation between X-ray and optical line emission falls into
three broad categories that are distinguished by surface brightness and
morphology: (1) regions of high surface brightness X-ray emission
distributed in long coherent arcs that are aligned tangentially with,
and lie interior to, the bright optical filaments; (2) localized
indentations in the shell, typically less than $20\arcmin$ 
in size,  where bright
X-ray emission occurs at large distances behind the projected edge of
the blast wave; and (3)
regions of low surface brightness, limb-brightened X-rays bounded on the
outside by faint, narrow Balmer-line filaments.

Examples of type (1) morphology are NGC~6992 in the northeast ($\alpha =
\hm{20}{56}, \delta = +32^\circ$, J2000)
and NGC~6960 on the western edge ($\hm{20}{46},+31^\circ$). 
Type (2) regions are found on the eastern edge ($\hm{20}{57}, +31^\circ$),
the southeast cloud ($\hm{20}{56}, +31^\circ$),
and the bright funnel-shaped section ($\hm{20}{50}, +32^\circ$) that is
located about $30\arcmin$ northeast of the optical emission region known as the
``carrot''  (Fesen, Blair, \&\ Kirshner 1982\markcite{FBK82}). 
These regions tend to be more turbulent and include the
most compact and highest surface brightness knots within the Cygnus Loop.
  Categories (1) and (2) occur where the shock has encountered
a greater than average column of gas, which has caused substantial
deceleration of the blast wave (Hester \& Cox 1986\markcite{HesCox86}; Hester et al.
 1994\markcite{HesRB94}; Graham et al. 1995\markcite{Gra95}; 
Levenson et al. 1996\markcite{Lev96}).  The difference
between regions (1) and (2) may be due to viewing geometry, the shape
of the cloud which is being encountered, and the time since enhanced
density was encountered by the shock.  The western edge
provides a vivid example of a blast wave-cloud interaction because this
region is viewed almost exactly edge-on, minimizing  the complications of
projection effects (Levenson et al. 1996\markcite{Lev96}).
The best examples of type (3) filaments occur at the extreme north
($\hm{20}{54}, +32^\circ$), to the south of the southeast knot, and to the north 
of NGC~6960.
These filaments show the location of the current projected edge of the
blast wave because Balmer-line filaments occur where previously
undisturbed atomic gas is shocked (Chevalier \&\ Raymond 1978\markcite{CR78};
Chevalier, Kirshner, \&\ Raymond 1980\markcite{CKR80}; Hester, Danielson, \& Raymond
1986\markcite{HesDR86}).
The X-rays do not completely overlap the optical emission but are offset
to the interior.  
We measure offsets of $5-20\arcsec$ between the
exterior edge of X-ray emission and the edge of the Balmer filaments;
offsets of $10\arcsec$ (equivalent to $10^{17}$ cm) are typical.
 The separation between the X-ray and optical emission
 is likely due to the time
required to heat the gas behind the shock front before
X-rays are emitted.  For a shock of velocity $v_s = 400 \kms$,
an offset of $20\arcsec$ implies a heating time of about 200 yr.
This is consistent with Coulomb collisions alone
equilibrating the postshock ion and electron temperatures 
(Spitzer 1962\markcite{Spitzer62}).  
If the blast wave has decelerated in dense gas where the Balmer filaments
are detected, the same separation from the X-ray edge implies that
a longer time has elapsed, so Coulomb equilibration alone is still 
a possible heating mechanism.  Whether Coulomb collisions or some
other mechanism heat the post-shock electrons, long equilibration
timescales are expected, as Laming et al. (1996\markcite{Lam96}) 
demonstrate in analysis of a non-radiative shock observed in SN 1006.

\section{A Global View}
\subsection{Cavity Explosion}
Various X-ray studies have successfully discussed individual
 regions of the Cygnus Loop in terms of blast wave-cloud collisions (e.g.
Hester \& Cox 1986\markcite{HesCox86}, Hester et al. 1994\markcite{HesRB94},
 Graham et al. 1995\markcite{Gra95}, Levenson et
al. 1996\markcite{Lev96}), and it is well established that large-scale collisions
fundamentally determine the morphology of portions of this 
supernova remnant.  However, the examination of particular regions does 
not lead to an understanding
of the global environment in which the SNR evolves.
Specifically, it is
important to determine whether the clouds that have been struck by the
blast wave are isolated and part of a random distribution or part of a
coherent structure, such as an H~II region cavity or wind blown bubble
formed by the supernova progenitor.  Previous global models of the
Cygnus Loop (e.g. Falle \&\ Garlick 1982\markcite{Fal82}; Tenorio-Tagle, Rozyczka, \&\
Yorke 1985\markcite{Ten85}) ignore the evidence for and effects of encounters with 
clouds and instead model a SNR breaking out of a molecular cloud.  
On the contrary, these X-ray observations emphasize that over large scales
the blast wave is running into dense material, not expanding into a 
contrasting lower density region.

With the data presented here, it is clear that the Cygnus Loop is the result of a cavity explosion.
The observations of the interior are consistent with blast wave propagation
through a rather smooth, low-density medium, and at the periphery, the
edge of the cavity has been encountered.
The enhancements in the ROSAT X-ray data show that the blast wave is currently 
interacting with
denser material---denser than the medium through which it previously
propagated---over 80\% of the perimeter that we have observed.  Some of
these enhancements are narrow, tangential filaments, while others are
the large structures that had been identified in earlier observations.
  Such extensive 
clumps of material cannot be characteristic of the pre-supernova ISM 
 within the current remnant and thus distinguish the cavity boundary.
We observe the cavity walls in these data! 
Only two extended sites, near $\alpha=\hm{20}{48}, \delta= +32^\circ 10\arcmin$,
 and
$\alpha=\hm{20}{57}, \delta= +30^\circ 30\arcmin$, show no evidence for blast
wave interaction with the cavity wall.

The cavity walls are nearly complete, consisting of a  neutral shell
as well as large clouds.
When the blast wave reaches the dense atomic shell, it decelerates. 
The Balmer-dominated filaments directly trace the motion of the
blast wave in the atomic gas of the shell.  
The X-ray signature of this category (3) morphology (cf. \S 3.1)
is enhancement at the SNR edge, where gas is piled up behind the
slowing shock.  Figures 8, 9, and 11 provide examples of this characteristic
X-ray profile in three different locations, where the enhancement occurs
at radii of $0.98 R_{snr}, 0.98 R_{snr}, {\rm \ and\ } 
0.96 R_{snr}$, respectively.
These enhancements are not due to the global expansion
of the blast wave in a uniform medium or evaporation in a uniform
distribution of clouds.  They are too slight, too limited
in radial extent, and misplaced relative to the Sedov and evaporative 
model predictions.
The Balmer filaments and associated X-rays are detected over half the
perimeter of the SNR, indicating that blast wave deceleration in atomic
material is common.  Some regions exhibit the type (3) X-ray signature
but the corresponding optical filaments are too faint to be detected.
  Although the blast wave
edge is not directly observed in these cases, the prevalence of 
category (3) and the observed X-ray morphology imply that the blast wave
is decelerating in the dense cavity wall in these locations, as well.

A radial profile across the northwestern edge
($\hm{20}{46}, +31^\circ 40\arcmin$) emphasizes the type (3)
X-ray limb-brightening 
and avoids confusion with emission due to interactions with large clumps
 (Fig. 11).
There is no extended bright optical emission due to radiative
shocks near the edge, and
this is not a particularly bright region of the X-ray map.
Two components are obvious, corresponding to two overlapping projections
of the blast wave that are not necessarily physically related. 
We neglect the interior component and 
model the exterior component very simply as a 
shock that strikes a density discontinuity and generates a 
reflected shock.
The Rankine-Hugoniot jump conditions determine the relationship 
among the flow variables in different regions.
The free parameters of the model are the density contrast and its 
initial location, from which the
density profile due the projected and reflected shocks as a function
of radius is calculated.  In the Cygnus Loop, 
Balmer-dominated filaments are typically observed
propagating through regions of $n\sim 1{\rm \ cm^{-3}}$ 
(e.g. Hester et al. 1994\markcite{HesRB94}, Raymond et al. 1983\markcite{Ray83}), 
while in the former H~II region
cavity $n \sim 0.1 {\rm \ cm^{-3}}$, so the density contrast is taken to be 10.
(This is likely to be an overestimate, however, because the brightest Balmer 
filaments were analyzed, and these are the ones that have evolved most
toward the radiative stage.)
The radius of the SNR blast wave 
is fixed at the location of the Balmer emission.
The interior edge of the X-ray enhancement marks the location of the
reflected shock.  We measure this to be at
 $R=0.94R_{snr}$, so the blast wave initially encountered the
wall at $R=0.96R_{snr}$ and is now at $R=1.0R_{snr}$.
  Similar to all the model calculations presented here,
the X-ray surface brightness profile is computed for the
spherically symmetric case (Fig. 11), assuming that
X-ray volume emissivity is proportional to density squared and is independent
of temperature.
This simple model reasonably 
matches the shape and location of the X-ray enhancement, although this
is not a fit to the data.
The model does not allow for 
heating time, so it predicts
 emission toward the edge of the SNR blast wave that we do not detect.
  Despite these approximations, however, the
simple model indicates the  scales of the parameters that cause
the X-ray enhancement at the edge where the blast wave has
run into the atomic cavity wall.

Several observed properties
 of the Cygnus Loop argue for a large-scale cavity explosion. The
first is the approximate spherical symmetry of the emission.  This symmetry
of the Cygnus Loop
is surprising in view of the rich X-ray morphology, which argues for an
inhomogeneous medium.  
If the inhomogeneity were in the form of a random distribution of 
clouds, then the clouds would have to be very small in order to account for
the fact that the 
observed gas produces few, if any, significant indentations
into the remnant
(McKee \&\ Cowie 1975\markcite{McKCow75}).  There are no entirely concave sections  
on scales greater than $20\arcmin$ (4 pc) and no evidence for thermal
evaporation of large clouds on the interior, for
example.
However, X-ray studies of individual
regions show that the clouds are quite extended along the
periphery of the remnant, with sizes of about $0\fdg5$ (7~pc).  
The clouds are preferentially aligned parallel to the blast wave.   In
this case, a  spherical remnant can be produced only
if neither the shape nor the distribution of the clouds is random.
To produce a
spherical remnant, the clouds in the vicinity of the Cygnus Loop must be
located at the edge of the current shell,
and they must be flattened along the shell. There is
also dynamical evidence that these are the first large clouds that
have been encountered by the blast wave: the blast wave-cloud
interactions to the northeast and the west are recent and approximately
the same age (Hester et al. 1994\markcite{HesRB94}, Levenson et al. 1996\markcite{Lev96}). 
This conclusion is consistent with that of Charles, Kahn, \&\ McKee (1985\markcite{Cha85}), who  
determined that data are
inconsistent with an explosion in a uniform distribution of clouds
based on Einstein-IPC observations adjacent to NGC~6960.
They suggested that the ISM within the current filamentary ring
lacked clouds before the supernova explosion.

\subsection{Progenitor Modification of the ISM}
The modification of the SNR environment by the pre-supernova star has
been discussed by McKee, Van Buren, \&
Lazareff (1984\markcite{McK84}) and Shull et al. (1985\markcite{Shull85}).
  If the progenitor star of the
Cygnus Loop SNR were an early B star, then the Lyman continuum radiation
from this main-sequence star created an H~II region. 
In the warm neutral interstellar medium with average density distribution
$n \approx 0.2 \exp(-z/220~{\rm pc})~{\rm cm}^{-3}$ 
(Kulkarni \&\ Heiles 1988\markcite{KulH88}),
a star at height $z=120$ pc emitting Lyman continuum photons at a rate
of $10^{46} {\rm s}^{-1}$ creates an H~II region with a  density
of about $0.1 {\rm \ cm^{-3}}$ and a radius of about 30 pc.
As denser clumps of gas 
within the H~II region are photoevaporated, adding more material
to the ionized interior, the radius of the H~II region  decreases.
Small clouds are ultimately destroyed by photoevaporation and larger clouds
are driven away by the rocket effect (Bertoldi \&\ McKee 1990\markcite{BerMcK90}).
The details of the photoevaporation depend on the velocity of the
progenitor star and on the properties of the clouds: temperature,
density, and magnetic field.  Using Bertoldi's (1989\markcite{Ber89}) results, one
can show that atomic clouds ($T \sim 10^2$\ K) in pressure balance
with the H~II region ($P/k \sim 2000 {\rm \ K \ cm^{-3}}$ for $n_{H~II} \sim
0.1 {\rm \ cm^{-3}}$) will be either destroyed or rocketed out of the H~II region.
Molecular clouds are more difficult to destroy by photoevaporation (Shull
 et al. 1985\markcite{Shull85}), but at the distance of the Cygnus Loop below the plane,
 very large molecular clouds are unlikely.  As a result of photoevaporation and
photodissociation, a
 dense atomic shell that contains most of the mass surrounds the H~II
region, which  is bounded by a recombination front.
A vigorous wind may aid formation of the
shell, although this is unlikely in the case of an early B star we
consider here. Once the progenitor leaves the main-sequence and becomes a red
giant, the H~II region begins to recombine, but the density is too low
for much recombination to occur before the star explodes.

When the progenitor of the Cygnus Loop
 became a supernova, the blast wave initially
expanded adiabatically through the diffuse cavity gas.  Eventually it hit
the walls of the cavity, which are comprised of the surrounding atomic shell
and the remains of clouds  that protrude into the H~II region.
NGC~6960 is a clear  example where the blast wave has encountered 
 dense molecular gas  (Levenson et al.
1996\markcite{Lev96}). The eastern edge and the southeast
 cloud may be additional examples of
elephant-trunk structures, such as those observed in 
 M~16 (e.g. Hester et al. 1996\markcite{Hes96}), protruding into the shell (Graham et
al. 1995\markcite{Gra95}).  The
existence of large clouds near the cavity edge is expected, for the
development of an H~II region tends to homogenize its interior, while
its action on the boundary is limited by intervening clouds along the
same line of sight and decreased ionizing flux at larger radii. 
The clouds currently 
at the edge were either rocketed out of the interior or were
originally located at the periphery of the H~II region.

We have considered a progenitor of spectral type later than B0 because
a stationary star of  earlier spectral type star would homogenize 
an H~II region to a radius of about 50 pc, while there are clearly
inhomogeneities within the present 20 pc extent of the SNR 
(McKee et al. 1984\markcite{McK84}; Charles et
al. 1985\markcite{Cha85}).  If a massive star is moving, however, the clouds situated
perpendicular to its path are not exposed to the Lyman continuum emission
for as long.  The rocket effect and photoevaporation are less effective on
these clouds, and the homogenizing radius across the direction of motion
is reduced.  Along the direction of motion, the distribution of the
processed interstellar
medium is anisotropic:  toward the original location of
the progenitor, gas has been exposed to ionizing radiation, 
so it is more homogeneous and less dense,
while the processing is limited in the direction that the
star moves, so the ISM is clumpier and less dense there.
The progenitor of the Cygnus Loop would have been later than B0, 
provided that it was not moving too rapidly; ($v_* \la$ a few \kms
so that surrounding clouds are exposed to the ionizing radiation for a time 
$R/v_* \ga 6\times 10^6$\ yr).  Alternatively, an earlier-type progenitor
is possible provided that it was moving predominantly along our line
of sight.  In this case, the projected edge of the cavity would still
be circular, yet have a smaller radius than one carved by a stationary
star.

\section{Conclusions}
The density inhomogeneities around the Cygnus Loop determine not only the
appearance of the SNR at optical and X-ray wavelengths, but also they
fundamentally alter the blast wave development.  The consequences are
significant beyond this single example.  Global models of
the interstellar medium that assume the clouds around supernova
remnants are insignificant or rely on statistical measures such as 
the number-diameter relation must be revised.

The high-resolution, global view of the Cygnus Loop these data provide 
resolves the apparent
paradox of the Cygnus Loop: a near-circular shape and evidence for
interaction with large-scale inhomogeneities in the interstellar medium.
The X-ray data make clear that the blast wave has recently encountered
density enhancements, which we identify as the walls of a cavity.  
The H~II region
of the progenitor star worked to homogenize a spherical cavity with 
ionizing radiation, and the blast wave initially propagated through this
cavity, maintaining its spherical symmetry.  Marking the edge of the 
cavity is a dense shell of atomic gas and large clumps of the processed
surrounding medium or formerly interior clouds that were rocketed out.
  Portions of the blast
wave have been decelerated in the smooth atomic shell, which
Balmer-dominated filaments and slightly limb-brightened X-ray profiles trace.
Other sections of the blast wave have swept up even larger column densities,
resulting in very bright X-rays and optical emission.  The observed
features of the Cygnus Loop are expected rather than surprising 
in the context of stellar evolution that preceded the supernova.

\acknowledgements
We thank the USRSDC for providing the ESAS software 
used to reduce the
X-ray data.  The optical images were obtained using the Prime Focus
Corrector on the 0.8-m telescope  
at McDonald Observatory in collaboration with L. Keller and M. J.
Richter.  The research of CFM is supported in part by NSF grant AST95-30480.

\newpage

%Fig. 1
\figcaption{ROSAT-HRI pointings superposed on the 
{\it Einstein\ } IPC image
of the Cygnus Loop (Seward 1990). %(Seward 1990\markcite{Sew90}).
Solid circles indicate fields that
are presented in this paper; dashed circles show unprocessed or proposed 
observations.}

%Fig. 2.
\figcaption{ROSAT-HRI count rate map of the Cygnus Loop.  
This image is a mosaic of
41 pointings in 26 positions.  Each observation has been background-subtracted
and exposure-corrected before being combined into this map, which has
$15\arcsec$ pixels.  It is scaled linearly from 0 to 0.15 counts ${\rm s^{-1} 
arcmin^{-2}}$.}

%Fig. 3.
\figcaption{Exposure map of combined ROSAT-HRI observations, including detector variations.}

%Fig. 4.
\figcaption{The boundaries of individual pointings are shown on the ROSAT-HRI
count rate mosaic of the Cygnus Loop.  The sharp, nearly circular
 edge of the supernova remnant
is real, not an artifact of the limited extent of the observations.}

%Fig. 5.
\figcaption{The northeast rim of the Cygnus Loop at $6\arcsec$ resolution.
{\it(a)}HRI data, which exhibits three distinct types of X-ray emission:
 long filaments of bright
emission, spots of extremely bright emission accompanied by indentations
in the X-ray shell, and fainter  limb-brightened filaments.
{\it (b)}The northeast rim of the Cygnus Loop in H$\alpha$, scaled linearly.  
The optical
image contains extended radiative filaments, small spots of bright
emission, and long, faint filaments due to Balmer-dominated shocks,
corresponding to the classes of X-ray emission.
{\it (c)}False-color composite of 
HRI (green) and H$\alpha$ (red)  observations, illustrating the exact
relation between the X-ray and optical emission.  The optical image
has been median-filtered to remove stars.}

%Fig. 6.
\figcaption{The southeast region of the Cygnus Loop at  $6\arcsec$ resolution.
{\it (a)}HRI, {\it (b)} H$\alpha$, and  {\it (c)} false-color composite,
as in Figure 5.}

%Fig. 7.
\figcaption{The northwest edge of the Cygnus Loop at  $6\arcsec$ resolution.
{\it (a)}HRI, {\it (b)} H$\alpha$, and  {\it (c)} false-color composite,
as in Figure 5.}

%Fig. 8.
\figcaption{HRI surface brightness as a function of radius across the 
bright northeastern rim of the Cygnus Loop.
This radial profile is an azimuthal average over $2\arcmin$.
Very bright emission appears projected to the interior of the SNR
and  extends over more than 12\% of  its radius.
While there is no significant X-ray emission at the edge of the SNR,
fainter X-ray limb-brightening appears to the interior, at 
$R=0.98R_{snr}$.  (Balmer-dominated optical filaments define the
edge of the SNR that is indicated in Figures 8, 9, and 11.)
This X-ray emission is clearly distinct from that
due to expansion of a blast wave in a uniform medium (cf. Fig. 10).}

%Fig. 9.
\figcaption{HRI surface brightness as a function of radius across the 
bright western rim of the Cygnus Loop, azimuthally averaged
over $2\arcmin$.  Extremely bright emission appears projected to
the interior of the SNR, at $R=0.92R_{snr}$.  Significant, though
less extreme, limb-brightening is also observed at $R=0.98R_{snr}$.}

%Fig. 10.
\figcaption{Model X-ray surface brightness profiles for Sedov SNR (solid line)
and evaporative cloud model of White \&\ Long (1991; dotted line) 
with their parameters $C=10$ and $\tau = 10$.  Both profiles have been
normalized to  a uniform level at the center of the supernova remnant.}

%Fig. 11
\figcaption{HRI surface brightness as a function of radius across the 
northwestern rim of the Cygnus Loop, azimuthally averaged
over $2\arcmin$.  Two separate components of the limb-brightened shell
are evident, near $R=0.96R_{snr}$ and $R=0.85R_{snr}$.
Both components are the result of interaction of the blast wave 
with the cavity wall, although the latter appears projected to the
interior of the SNR.
The outer peak is modeled as the result of a density
contrast and reflected shock (dotted line).  In this
model, the blast encounters a density contrast of 10 at $R=0.96R_{snr}$
and the reflected shock has progressed to $R=0.94R_{snr}$  when the
forward shock has reached $R=1.0R_{snr}$}


\begin{references}
\reference{Asch88}Aschenbach, B. 1988, \ao, 27, 1404
\reference{Ber89}Bertoldi, F. 1989, \apj, 346, 735
\reference{BerMcK90}Bertoldi, F., \&\ McKee, C. F. 1990, \apj, 354, 529
\reference{Cha85}Charles, P. A., Kahn, S. M., \&\ McKee, C. F. 1985, 
 \apj, 295, 456
\reference{CKR80}Chevalier, R. A., Kirshner, R. P., \&\ Raymond, J. C. 1980,
 \apj, 235, 186
\reference{CR78}Chevalier, R. A., \&\ Raymond, J. C. 1978, \apj, 225, L27
\reference{Cow81}Cowie, L. L., McKee, C. F., \&\ Ostriker, J. P. 1981,
 \apj, 247, 908
\reference{DeN75}DeNoyer, L. K. 1975, \apj, 196, 479
\reference{Fal82}Falle, S. A. E. G., \&\ Garlick, A. R. 1982, \mnras, 
 201, 635
\reference{FBK82}Fesen, R. A., Blair, W. P., \&\ Kirshner, R. P. 1982,
 \apj, 262, 171
\reference{Fes92}Fesen, R. A., Kwitter, K. B., \&\ Downes, R. A. 1992,
 \aj, 104, 719
\reference{Gra96}Graham, J. R. \&\ Aschenbach, B. 1996,
 in ROSAT Newsletter \#13, Ed. S. L. Snowden, 6
\reference{Gra95}Graham, J. R., Levenson, N. A., Hester, J. J., 
 Raymond, J. C.,  \&\ Petre, R. 1995, \apj, 444, 787
\reference{Ham83}Hamilton, A. J. S., Sarazin, C. L., \&\  Chevalier, R. A. 
 1983, \apjs, 51, 115
\reference{Hes96}Hester et al. 1996, \aj, 111, 2349
\reference{HesCox86}Hester, J. J., \&\ Cox, D. P. 1986, \apj, 300, 675
\reference{HesDR86}Hester, J. J., Danielson, G. E., \& Raymond, J. C.
1986, \apj, 303, L17
\reference{HesRB94}Hester, J. J., Raymond, J. C. \&\ Blair, W. P. 1994,
 \apj, 420, 721
\reference{Klein94}Klein, R. I., McKee, C. F., \& Colella, P. 1994, \apj, 420, 213
\reference{Klein95}Klein, R. I., McKee, C. F., \& Woods, D. T. 1995,
 in The Physics of the Interstellar Medium and Intergalactic Medium,
 ed. A. Ferrara, C. F. McKee, C. Heiles, and P. R. Shapiro 
 (San Francisco: ASP), 366
\reference{Ku84}Ku, W. H.-M., Kahn, S. M., Pisarski, R. \&\ Long, K. S. 1984,
 \apj, 278, 615
\reference{KulH88}Kulkarni, S. R., \&\ Heiles, C. 1988, in Galactic and
 Extragalactic Radio Astronomy, 2nd ed.,  ed. G. L. Verschuur and 
 K. I. Kellermann, (New York: Springer-Verlag), 95
\reference{Lam96}Laming, J. M., Raymond, J. C., McLaughlin, B. M., \&\
 Blair, W. P. 1996, \apj, 472, 267
\reference{Lev96}Levenson, N. A., Graham, J. R.,  Hester, J. J.,
 \&\ Petre, R. 1996, \apj, 468, 323 
\reference{Lev97}Levenson, N. A., Graham, J. R.,  Keller, L., \&\ 
 Richter, M. J. 1997, in preparation
\reference{McK87}McKee, C. F. 1987, in Spectroscopy of Astrophysical Plasmas, 
 ed. A. Dalgarno and D. Layzer, (New York: Cambridge University Press.), 226
\reference{McKCow75}McKee, C. F. \& Cowie, L. L. 1975, \apj, 195, 715
\reference{McK84}McKee, C. F., Van Buren, D. \&\ Lazareff, B. 1984,
 \apj, 278, L115
\reference{Mink58}Minkowski, R. 1958, Rev.~Mod.~Phys, 30, 1048
\reference{Ray83}Raymond, J. C., Blair, W. P., Fesen, R. A., \&\ Gull, T. R. 
 1983, \apj, 275, 636
\reference{Sco77}Scoville, N. Z., Irvine, W. M., Wannier, P. G., \&\ Predmore,
 C. R. 1977, \apj, 216, 320
\reference{Sedov93}Sedov, L. I. 1993, Similarity and Dimensional Methods 
 in Mechanics, 10th ed., (Boca Raton: CRC Press)  
\reference{Sew90}Seward, F. D. 1990, \apjs, 73, 781
\reference{Shull85}Shull, P., Jr., Dyson, J. E., Kahn, F. D., \&\ West, K. A.
 1985, \mnras, 212, 799
\reference{Sno94}Snowden, S. L., McCammon, D., Burrows, D. N., \&\ Mendenhall,
 J. A. 1994, \apj,  424, 714
\reference{Spitzer62}Spitzer, L., Jr. 1962, Physics of Fully Ionized Gases,
 2nd ed., (New York: Wiley)
\reference{Ten85}Tenorio-Tagle, G., Rozyczka, M., and Yorke, H. W., 1985,
 \aap, 148, 52
\reference{WhL91}White, R. L., \&\ Long, K. S. 1991, \apj, 373, 543
\reference{Zom90}Zombeck, M. V., et al. 1990, \procspie, 1344, 267

\end{references}
\end{document}